\documentclass[aps,amsmath,amssymb]{revtex4} 

\usepackage{graphicx}
\usepackage{dcolumn}
\usepackage{bm}
\usepackage{xcolor}

\begin{document}

\bibliographystyle{apsrev} 

\title{Investigation of the unidirectional spin heat conveyer effect in a 200nm thin Yttrium Iron Garnet film}

\author{O. Wid$^{1}$}%
\author{J. Bauer$^{2}$}%
\author{A. M\"{u}ller$^{1}$}%
\author{O. Breitenstein$^{2}$}%
\author{S. S. P. Parkin$^{2}$}%
\author{G. Schmidt*$^{1,3}$}%

\affiliation{%
$^1$Institut f\"{u}r Physik, Martin-Luther-Universit\"{a}t Halle-Wittenberg, Halle, 06120, Germany \\
$^2$Max Planck Institute of Microstructure Physics, Halle, 06120, Germany \\
$^3$IZM, Martin-Luther-Universit\"{a}t Halle-Wittenberg, Halle, 06120, Germany \\
$*$Address for correspondence: georg.schmidt@physik.uni-halle.de\\
}
\begin{abstract}
We have investigated the unidirectional spin wave  heat conveyer effect in sub-micron thick yttrium iron garnet (YIG) films using lock-in thermography (LIT). Although the effect is small in thin layers this technique allows us to observe asymmetric heat transport by magnons which leads to asymmetric temperature profiles differing by several mK on both sides of the exciting antenna, respectively. Comparison of Damon-Eshbach and backward volume modes shows that the unidirectional heat flow is indeed due to non-reciprocal spin-waves. Because of the finite linewidth, small asymmetries can still be observed when only the uniform mode of ferromagnetic resonance is excited. The latter is of extreme importance for example when measuring the inverse spin-Hall effect because the temperature differences can result in thermovoltages at the contacts. Because of the non-reciprocity these thermovoltages reverse their sign with a reversal of the magnetic field which is typically deemed the signature of the inverse spin-Hall voltage.

\end{abstract}

\flushbottom
\maketitle
%
%
\thispagestyle{empty}

\section*{Introduction}

In 2013 An \textit{et al.} have shown that by the excitation of nonreciprocal spin waves, so-called Damon-Eshbach modes (DEM), in a 400 $\mu$m thick Yttrium Iron Garnet (YIG) crystal heat can be transported in the absence of a temperature gradient \cite{An2013}. The direction of the heat flow can be reversed by reversing the applied magnetic field. In their experiments they observed two different effects, both based on the non-reciprocity of the DEM, however, with different consequences. On one hand the asymmetric excitation and propagation of the spin-waves leads to an asymmetric temperature profile which is dominated by the energy loss of the spin waves. These losses are highest at the point of excitation and decrease with increasing distance from this point because the amplitude of the spin-waves decreases. Nonetheless, the energy transport still occurs from the point of higher temperature towards lower temperatures, however, with a certain asymmetry with respect to the source because of the asymmetric propagation of the spin-waves. In an infinite sample no further effects would be observed. If on the other hand the sample is small enough for the spin-waves to actually reach the edge of the sample an additional effect occurs. In this case the spin waves, due to the non-reciprocity cannot be reflected and thus deposit all remaining energy as heat at the boundary. This results in an increase in temperature realizing actually heat transfer by magnons into a warmer region and thus along the temperature gradient as stated by An \textit{et al.} (normally heat transport is against the temperature gradient).
In 2015 more detailed theoretical descriptions have been published based on a phenomenological theory \cite{Adachi2015} and on micromagnetic simulations \cite{Perez2015}. The measurements reported so far were performed with an infrared camera on YIG films with thicknesses ranging from a few micrometers to hundreds of micrometers.\\
Current research in magnonics concentrates more and more on thin ferromagnetic films and often uses the measurement of the inverse spin-Hall-effect which relies on the measurement of very small voltages. It is thus quite important to know whether a similar scenario can also occur in ultra thin films ultimately leading to temperature gradients which can lead to thermovoltages whose sign might then also depend on the magnetic field.

The DEMs used in the experiment are nonreciprocal surface spin waves \cite{Eshbach1960}. They propagate in opposite directions ($\vec{k}$ and $-\vec{k}$) at the top and bottom surface of the layer, respectively \cite{An2013, Adachi2015}. The precessional amplitude of the DEM is maximum at the surface and decays exponentially inside the film \cite{Serga2010}. In order to cause unidirectional heat flow the population of $\vec{k}$ and $-\vec{k}$ must be different. In a thick sample this condition is realized by exciting the spin waves using a microwave antenna \cite{An2013} which is in contact with one side of the layer. In this case the spin waves at the surface close to the antenna are excited more strongly then at the other side resulting in a net spin wave current in one direction.

For thin films with a thickness of a few hundred nm the situation is quite different. The distribution of the precessional amplitude across the film thickness is almost uniform \cite{Serga2010} as well as the excitation by the antenna at the top and bottom surface described above. Nevertheless, the population and propagation of the $\vec{k}$ and $-\vec{k}$ vectors can still be different. The bottom surface of a thin YIG film which is typically grown on gadolinium gallium garnet (GGG) is in contact with the paramagnetic substrate while the top surface is in contact to air. Due to this fact the damping of the spin waves at the top and bottom surface can be different, leading to increased damping of the spin waves propagating in one direction. Also the excitation by the antenna can be nonuniform for thin films. If a waveguide is used for excitation in the Damon-Eshbach geometry the in-plane and the out-of-plane component of the microwave magnetic field can both excite spin waves. The interference of these waves is destructive for k-vectors in one direction perpendicular to the antenna and constructive for the opposite direction \cite{Serga2010, Schneider2008}. Moreover in thin films the damping is higher and the spin-waves are less likely to reach the boundary of the sample so we will not expect the heat pile up that An et al. were able to observe.\\
Thus in thin YIG films the unidirectional spin wave heat conveyer effect is expected to be present in terms of an asymmetric temperature profile, however, it will be small in magnitude. Steady-state infrared cameras used in the measurements reported so far are most likely not sensitive enough to detect the effect in thin films. It is, however, possible to use lock-in thermography (LIT), which is well established for failure analysis in integrated circuits \cite{Breitenstein2000} and the characterization of solar cells \cite{Breitenstein2010, Bauer2009}. The LIT is a dynamic method, which detects temperature modulation in infrared images similar to electrical measurements using a lock-in amplifier. With this technique the difference in temperature between an excited and a non excited state is imaged. Temperature differences as small as 100 $\mu$K can be resolved which is sensitive enough for the small effects described above as we will show later. Details about the used LIT system can be found in the Methods Section.

\section*{Experimental Setup}
\begin{figure}[ht]
\centering
\includegraphics[width=.5\linewidth]{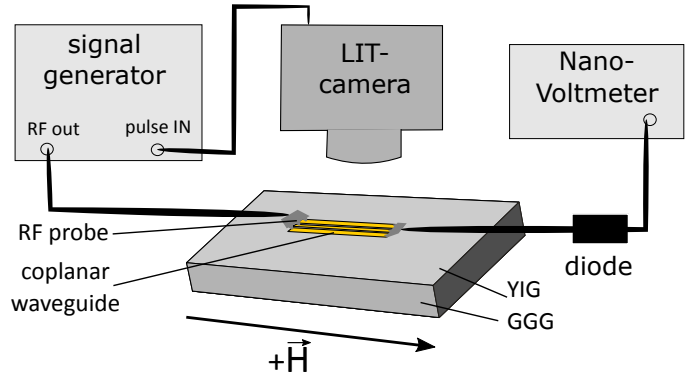}
\caption{Experimental setup for the FMR and the LIT measurements. The FMR measurements are performed by applying a continuous microwave with a constant frequency and sweeping the magnetic field. For the LIT measurements the microwave has to be pulsed with the lock-in frequency, which is provided by the camera.}
\label{fig:setup}
\end{figure}

A sketch of our experimental setup is shown in Figure \ref{fig:setup}. We use a 200 nm high quality YIG film on Gadolinium Gallium Garnet (GGG) substrate grown by liquid phase epitaxy. The damping in ferromagnetic resonance (FMR) $\alpha$ of the layer is smaller than $1\times10^{-4}$. To excite spin waves we use a coplanar waveguide (CPW) as an antenna, which is fabricated on top of the sample (size of the sample: 5 mm $\times$ 8 mm). Details about the fabrication and dimensions of the CPW can be found in the Methods Section.\\
In order to investigate the spin-wave spectrum FMR measurements are performed. The corresponding experimental setup is shown in Fig. \ref{fig:setup}. Measurements are done by applying a continuous microwave to the antenna and measuring the transmitted signal using a diode and a nanovoltmeter while sweeping the magnetic field.

For the LIT measurement the camera is placed above the sample and the lock-in reference frequency provided by the camera is used to pulse the microwave excitation.

\section*{Ferromagnetic resonance (FMR) measurements}
\begin{figure}[ht]
\centering
\includegraphics[width=.5\linewidth]{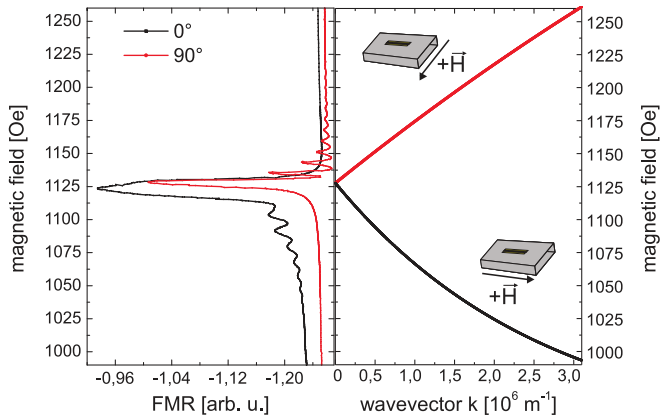}
\caption{Result of the FMR measurement at 5 GHZ and the calculated dispersion relation for a 200 nm thin YIG film using the following values: saturation magnetization $4\pi M_0$=1700 Oe, gyromagnetic ratio $\gamma$=2.8 $\frac{MHz}{Oe}$. }
\label{fig:dispersion}
\end{figure}

FMR is measured at a constant frequency of 5 GHz and an excitation power of 1 dBm while sweeping the external magnetic field from 990 Oe to 1260 Oe. The external field is always aligned in the plane of the layer and is either kept parallel to the antenna to excite the Damon-Eshbach mode with $\vec{k}\perp\vec{M}$ (DEM-geometry) or perpendicular to the antenna to excite the backward volume mode with $\vec{k}\|\vec{M}$ (BVM-geometry). The result of the two respective FMR measurements are shown in Figure \ref{fig:dispersion} together with the calculated dispersion relation for an in-plane magnetized 200 nm YIG film. These dispersion curves for dipolar spin waves have been calculated using the equations given in \cite{Serga2010}. For $k=0$ the uniform mode is excited at a field of 1128 Oe. For $k\neq0$ the dispersion relation exhibits two branches, one for the DEM at lower fields and one for the BVM at higher fields. Comparing the FMR spectra to the dispersion relation we can see that for both geometries the uniform mode is excited. In the DEM geometry we also observe a signal at lower $\vec{H}$ while for the BVM-geometry resonances at higher $\vec{H}$ appear in good agreement with the calculation. We do, however, not observe the expected continuous spin-wave spectrum but several resonance lines. These lines appear because the geometry of the waveguide favours certain k-vectors corresponding to a fundamental k-value and integer multiples with decreasing amplitude. It should be noted that the DEM with the highest intensity overlaps with the uniform mode and is not visible as a separate peak.

\section*{Lock-in thermography measurements}

\subsection*{Damon-Eshbach geometry}
\begin{figure}[ht]
\centering
\includegraphics[width=\linewidth]{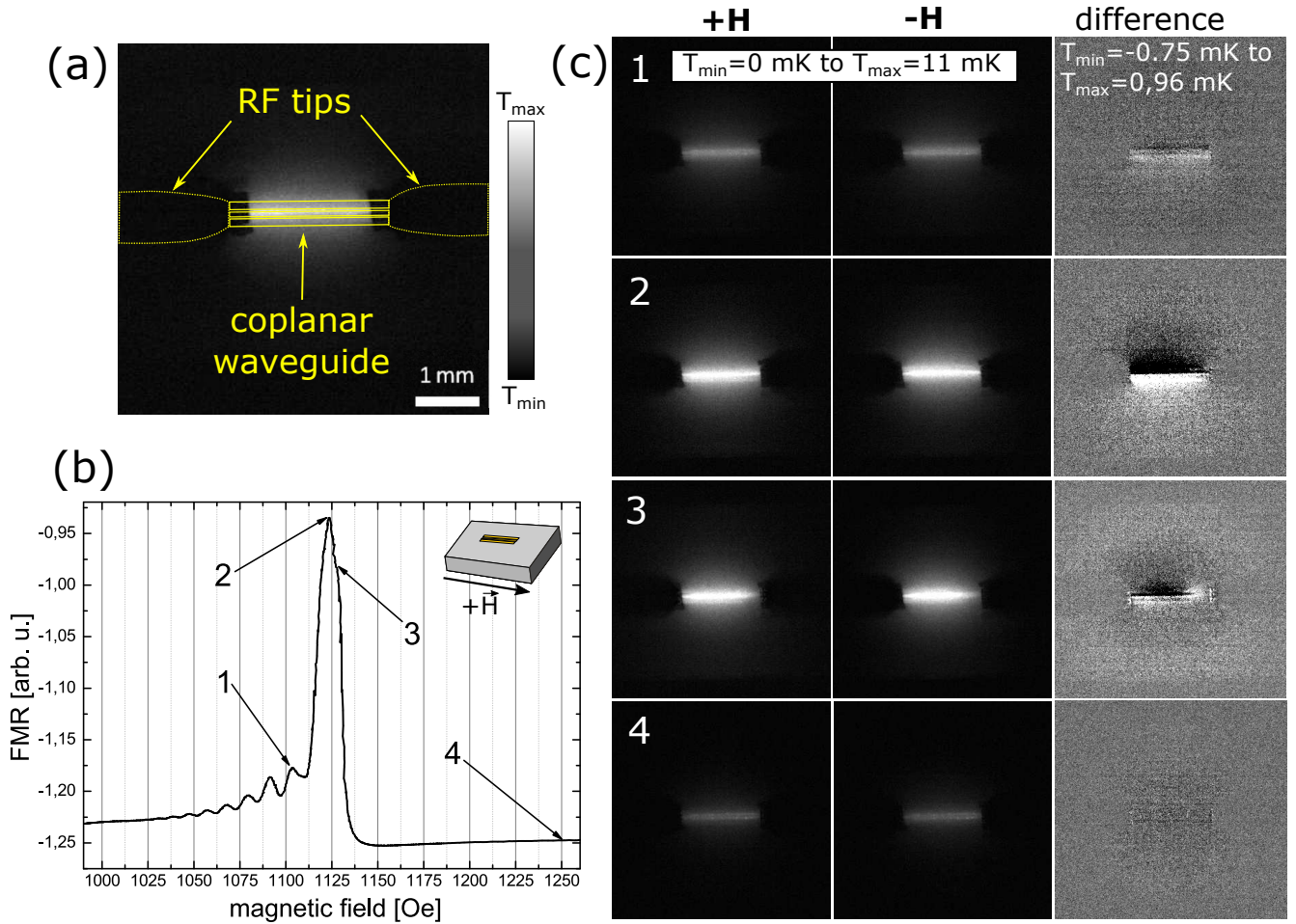}
\caption{(a) LIT amplitude image with a sketch of the position of the CPW and the rf tips. (b) FMR measurement in the Damon-Eshbach geometry (c) LIT measurements when the magnetic fields 1, 2, 3, 4 and the corresponding negative values are applied (left and center). Black corresponds to no increase in temperature while white indicates an increase of $11 mK$. The calculated difference (right) shows the expected effect. It should be noted that in the difference images the gray scale has been changed for better visibility. The range is now between -0.75mK (black) or lower and 0.96 mK (white) or higher.}
\label{fig:dem}
\end{figure}

For the lock-in thermography measurements we use the knowledge obtained from the ferromagnetic resonance spectra. First the DEM-geometry is investigated. Using different field values we excite different parts of the spin wave spectrum and take images with the camera. Figure \ref{fig:dem}, (a) shows an amplitude image taken by the LIT camera at a magnetic field of approx. 1120 Oe including a sketch of the rf tips connected to the coplanar waveguide. Bright color in the LIT images corresponds to higher temperature than dark color. These measurements are repeated at different magnetic fields which are marked in Figure \ref{fig:dem},(b). These values correspond to the maximum DEM peak which is not overlapping with the uniform mode (1), the maximum DEM peak overlapping with the uniform mode (2), the uniform mode itself (3), and a position completely off resonance where no excitation is expected at all (4), respectively. All measurements are repeated after reversing the direction of the magnetic field.

For the interpretation of the images it is important to first understand what is actually visible. As a typical lock-in technique the LIT is able to eliminate background signals. In this case the absolute temperature is not measured. The difference between the state without and with excitation, however, is measured with high accuracy. The temperature given by the gray scale is thus the increase in temperature arising from the excitation. We do not expect to measure negative values because no active cooling is expected in our case. The background which is eliminated by the lock-in technique corresponds to room temperature. The accuracy of the measurements has also been determined by comparing different measurements and we can show that the error is typically $0.15\,mK$ or less. Error bars are thus omitted when temperature profiles are displayed in graphs. It should be noted that the local emissivity can lead to deviations of $\pm 5- \pm 10\%$ which can locally distort the temperature profile. However, as we will see later, we obtain our results mainly from the difference between two measurements in which the systematic error of the emissivity is eliminated.

In order to correctly evaluate the data it is also necessary to understand and eliminate possible side effects and artifacts. First the influence of the antenna should be discussed. As also outlined in the Methods Section the whole sample surface is covered by black ink to ensure uniform emission properties for heat radiation eliminating the low radiation efficiency of a blank metal surface. The question remains, however, whether in the areas of metallization the images really show the temperature of the YIG or of the metal and whether any heating really stems from the resonance in the oxide layer rather than from losses in the coplanar waveguide. The latter can easily be excluded by comparing pictures off-resonance and on-resonance which indicate that the heating by the antenna observed off-resonance is much smaller than the heating by the resonance in the YIG. In addition, this heating does not change during a field reversal and can thus be eliminated by calculating the difference between images with opposing magnetic fields. Visibility of the YIG temperature through the metal is also guaranteed because the thickness of the metal is only a few hundred nm and the materials of the coplanar waveguide (Ag and Au) have very high heat conductivity. At the modulation frequency of 1 Hz the top of the antenna can always be considered at the same temperature as the YIG surface underneath. Lateral heat conduction in the thin metal film, however is much lower and will only slightly smear any temperature profile originating from the spin waves. The same holds for heat diffusion in the YIG. Anyway, both, heat diffusion in metal and YIG are again independent from the magnetic field and can thus be eliminated by taking the difference of two images with opposite field directions. They will, however, slightly reduce the effect that we intend to observe.

Fig. \ref{fig:dem}, (c), 1 shows the image taken for one H-field direction at the smaller DEM peak (position 1, Fig. \ref{fig:dem},(b)). The image itself does not allow us to directly observe any asymmetry in the temperature profile and the image taken after field reversal looks quasi-identical. We now calculate the difference between the two pictures. The grey scale of the pixels now no longer corresponds to an increase in temperature due to excitation but to a temperature difference between two excited states. Negative values can occur, however they are no indication of cooling but just mean that the corresponding local temperature on the subtracted image is higher than on the image from which is subtracted.
For this first case the difference image already shows a small asymmetry between the two sides of the antenna. Moving to the maximum intensity of the DEM (position 2, Fig. \ref{fig:dem},(b), (c)) not only shows a much bigger increase in temperature but here also the difference image clearly shows an asymmetric temperature profile extending over several mm across and beyond the antenna.

\begin{figure}[ht]
\centering
\includegraphics[width=.5\linewidth]{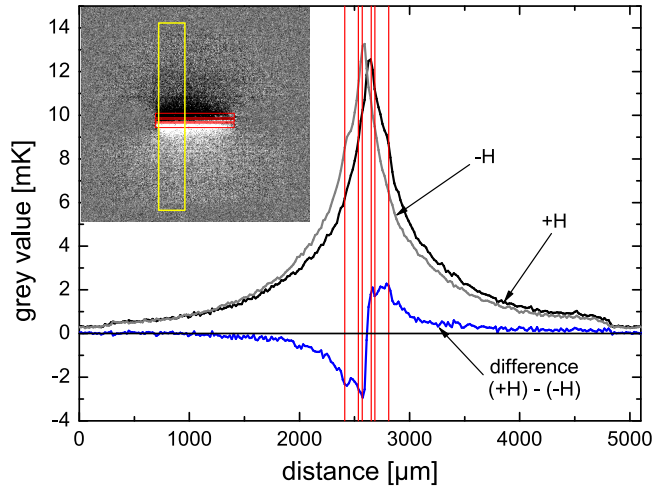}
\caption{Temperature profile plotted for the yellow marked region. Red lines show the position of the CPW.}
\label{fig:profile}
\end{figure}

Figure \ref{fig:profile} shows a temperature line profile obtained by line-wise averaging the data inside the yellow region shown in the inserted image for positive (black curve) and negative (grey curve) H-field respectively. For the diagram the grey value is converted to the temperature in mK. Already here we can observe that decrease in temperature away from the antenna is faster on one side than on the other leading to an asymmetric temperature profile. This asymmetry is reversed when the magnetic field changes sign. For better visibility we also plot the difference between the two graphs showing that the maximum temperature difference between both sides of the antenna is as big as 5.2 mK and decays to zero far away from the antenna as expected from theory. Negative temperatures only appear because the difference has been calculated. It should be emphasized  that the fact that we can observe the asymmetrical temperature distribution also outside of the region of the antenna (red lines in Fig. \ref{fig:profile}) clearly shows that the effect indeed originates from propagating spin waves.

For the uniform mode at line position 3 (Fig. \ref{fig:dem},(b) (c), 3) the heating should be symmetric and independent from the direction of the magnetic field. Nevertheless, we still observe a small asymmetry which results from the overlap with the DEM at position 2. This effect will later be discussed because it is highly relevant for measurements of the inverse spin-Hall effect (ISHE) \cite{Qiu2015}. As expected we only observe little heating and a homogeneous temperature distribution for the off-resonance measurement (Fig. \ref{fig:dem}, (b), (c), 4).

\subsection*{Backward Volume geometry}
\begin{figure}[ht]
\centering
\includegraphics[width=\linewidth]{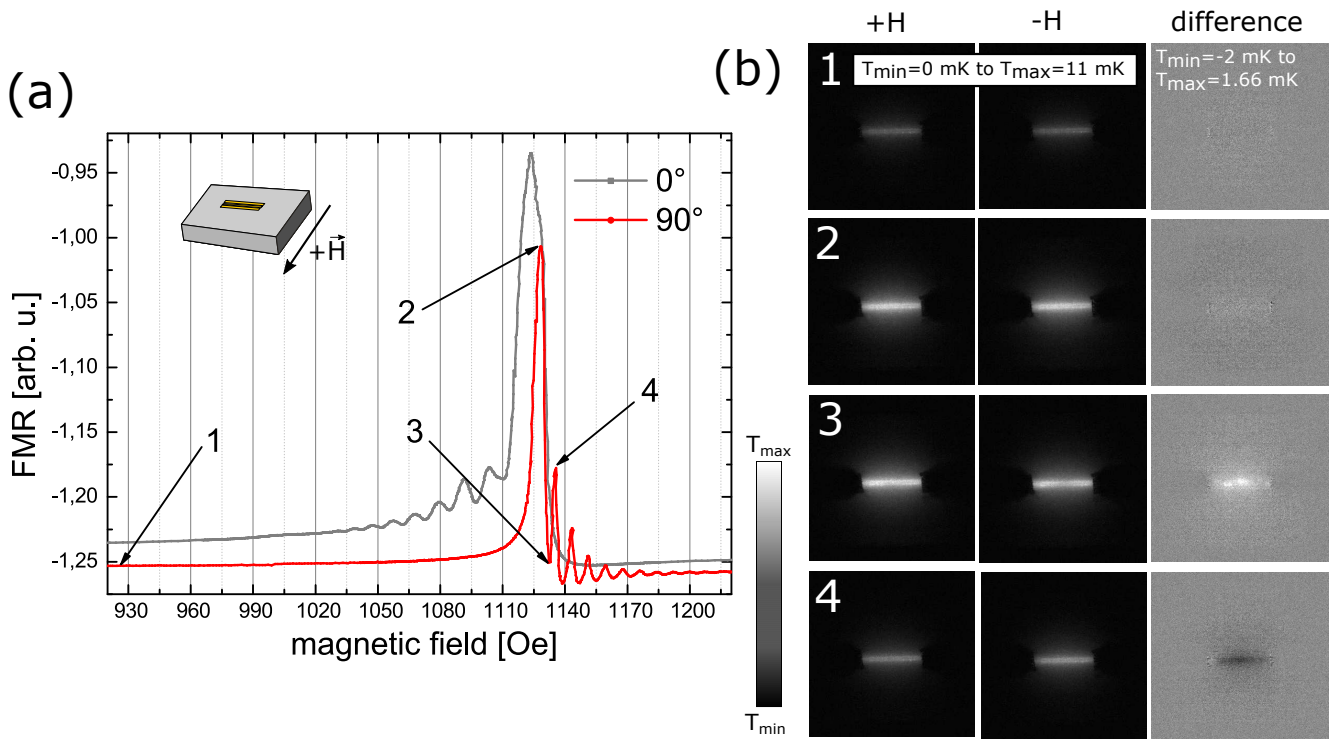}
\caption{(a) FMR measurement in the Backward Volume geometry ($90^\circ$, red curve) compared to the FMR measurement in the Damon-Eshbach geometry ($0^\circ$, grey curve). The peaks of the BVM appear much sharper than for the DEM. (b) LIT images for the magnetic field positions 1 to 4.}
\label{fig:bvm}
\end{figure}

For the BVM geometry a similar set of measurements is done as for DEM, now sampling the field range of the backward volume modes including the uniform resonance mode. In the Backward Volume geometry no non-reciprocal spin-waves can be excited, so that no unidirectional heat transfer should be observed. This is confirmed by Figure \ref{fig:bvm} which displays the spin wave spectrum together with the LIT images again for four different measurements. For the off-resonance case (position 1) and for the uniform mode (line 2) symmetric temperature profiles with a maximum at the antenna appear and the difference images indicate no difference at all. For position 3 and 4 the difference is non-zero, however, we can identify this as an artifact. Although the difference is finite, it is not asymmetric with respect to the antenna. Indeed it does not originate from non-reciprocity of the spin waves but from an absolute increase in heating for one field direction. The origin of this difference is the sharpness of the resonance lines in the BVM-geometry. The pictures are taken at fixed field positions and it can actually happen that upon field reversal a slightly different field value is applied. Even a difference of 0.5 Oe (corresponding to a relative error of $500\,ppm$) can lead to a sizeable difference in heating, explaining the observed effect.

\section*{Discussion}\label{discussion}

Our results show that even for thin YIG films the unidirectional spin wave heat conveyer effect can be observed although we do not see the heat pile-up shown by An and coworkers due smaller range over which the spin-waves travel in our samples. Although the temperature differences and the lateral extent of the effect are smaller than for thick YIG films, lock-in thermography allows us to clearly identify the effect. The magnitude of the signal that is observed indicates that it should also be present in even thinner layers which nowadays can also be obtained with very low damping (for example \cite{Kelly2013, Chang2014, Hauser2016} ) and may possibly be detected using longer averaging times. The fact that this effect needs to be considered also in thin films has important consequences for other research areas. In our experiments we observe temperature differences of several mK over a range of several mm. The DEM geometry of excitation is also the one which is used when the inverse ISHE is measured. Looking at the details of such measurements we find that the heat-conveyer effect creates a temperature gradient along the direction in which the DC inverse spin-Hall voltage is measured. This temperature gradient now leads to thermovoltages which, in contrast to other typical thermoelectric artifacts, exhibit the same symmetry with respect to the magnetic field as the ISHE. They can thus not be ruled out by comparing measurements at opposite field directions. Furthermore at least in our experiment the finite line width prevents us from completely separating the uniform mode from the DEM which would be necessary to make sure that no temperature gradient is created. This shows that great care must be taken in order to clearly discern ISHE and possible thermovoltages, especially when the ISHE is very small. In particular when organic materials are used the possibly large Seebeck coefficient can result in voltages as high as those observed for the ISHE. The conducting Polymer  poly(3,4-ethylenedioxythiophene) polystyrene sulfonate (PEDOT:PSS) which is often applied in organic electronics can for example exhibit a Seebeck coefficient of $160\,\mu K$ or higher \cite{Massonnet2014, Bubnova2014} resulting in a thermovoltage of 160 nV for a temperature difference of 1 mK which we even achieve in our experiment. For thick YIG layers where the heat conveyer effect is much more pronounced the expected thermovoltages can thus be much bigger.

\section*{Methods}\label{methods}
\subsection*{Fabrication of the CPW}
The CPW is fabricated on top of the YIG film using electron beam lithography, metal deposition (10nm Ti / 250nm Ag / 50 nm Au) and lift-off. The dimensions of the waveguide are as follows: length = 2.9 mm, width of the signal line = 80 $\mu$m, width of the ground planes = 125 $\mu$m, distance between signal line and ground planes = 35 $\mu$m.

\subsection*{Ferromagnetic resonance (FMR) measurement}
All FMR measurements shown in this paper have been performed using a microwave frequency of 5 GHz and a excitation power of -1 dBm, by sweeping the magnetic field. The external magnetic field is applied by an electromagnet which can be rotated around the sample. The microwave is provided by a RHODE\&SCHWARZ, SMF 100A signal generator, which is connected via rf probes (Cascade Microtech) to the CPW on top of the YIG sample. While sweeping the magnetic field at a constant frequency we measure the absorption using a Schottky diode and an Agilent 34420A nanovoltmeter.

\subsection*{Lock-in thermography technique}
For the LIT experiments we used an InfraTec PV-LIT system, which works with an InSb detector having a resolution of 640 x 512 pixels \cite{infratec}. To perform the LIT measurement the microwave power is pulsed at the lock-in frequency supplied by the camera. For all LIT measurements shown in this paper we use an acquisition time of 5 minutes and a lock-in frequency of 1 Hz. The surface of the sample is blackened with ink to achieve a better and uniform infrared emissivity. In a LIT experiment the heat sources in a device are modulated or pulsed at a lock-in frequency lying well below the frame rate of the infrared camera (here 200 Hz). The aquired images are evaluated synchronously to the heat pulses to detect the temperature modulation. Both the in-phase and the out-of-phase modulations are detected and can be converted into an amplitude and a phase signal. This is done on-line for each pixel. The result is equivalent to connecting each pixel to a two-phase lock-in amplifier \cite{Breitenstein2000, Breitenstein2010, Bauer2009}.

\section*{Acknowledgements}

This work was supported by the Deutsche Forschungsgemeinschaft in the SFB\,762 . We thank Georg Woltersdorf for fruitful discussion.




\end{document}